\documentclass[twocolumn,reprint,showpacs,pre]{revtex4-1}
\usepackage[T1]{fontenc}
\usepackage[utf8]{inputenc}
\usepackage{amsmath,amsfonts,amssymb}
\usepackage{graphicx}
\newcommand\EE[1][1]{\mathrm{E}\sb{#1}}

\renewcommand\d{\,\mathrm{d}}

\newcommand\e{\mathrm{e}}

\renewcommand\i{\mathrm{i}}

\newcommand\vk{\mathbf k}

\renewcommand\Re{\mathrm {Re}}

\newcommand\uu{\hat{\mathbf u}}

\renewcommand\vr{\mathbf r}

\renewcommand\S[1]{\mathcal{S}\sp{#1}}

\newcommand\TF[1]{\widetilde{#1}}

\newcommand\TL[1]{\widehat{#1}}

\renewcommand\TH[1]{\check{#1}}

\newcommand\TFL[1]{\overline{#1}}

\newcommand\derp[3][]{\partial\sp{#1}\sb{#3}{#2}}

\newcommand\Gei{\mathit{g}}

\newcommand\Gri{\mathit{G}}

\newcommand\Gru{\mathbb{G}}

\newcommand\Grua{\Gru}

\newcommand\fei[1][]{#1{\Gei}\sb0}

\newcommand\fri[1][]{#1{\Gri}\sb0}

\newcommand\fru[1][]{#1{\Gru}\sb0}

\newcommand\gei[1][]{#1{\Gei}}

\newcommand\gri[1][]{#1{\Gri}}

\newcommand\gru[1][]{#1{\Gru}}

\newcommand\geistst[1][]{#1{\Gei}\stst\sb\infty}

\newcommand\gsristst[1][]{#1{\Gri}\stst\sb1}

\newcommand\gristst[1][]{#1{\Gri}\stst\sb\infty}

\newcommand\gsru[1][]{#1{\Gru}\sb1}

\newcommand\gdrustst[1][]{#1{\Gru}\stst\sb2}

\newcommand\grustst[1][]{#1{\Gru}\stst\sb\infty}

\newcommand\grua[1][]{#1{\Grua}}

\newcommand\stst{\sp{\mathrm{ss}}}

\begin{document}
\title[Radiative transfer in two dimensions]%
{Space-time domain velocity distributions in 
isotropic radiative transfer in two dimensions} 
\author{Vincent Rossetto
\footnote{E-mail:\texttt{vincent.rossetto@grenoble.cnrs.fr}}}
\address{
    Universit\'e Grenoble Alpes / CNRS\\
    Laboratoire de physique et mod\'elisation des milieux condens\'es\\
    Maison des Magist\`eres - CNRS, BP 166\\
    25, avenue des Martyrs, 38042 Grenoble CEDEX, France
}

\begin{abstract}
We compute the exact solutions of the radiative transfer equation in two
dimensions for isotropic scattering. The intensity and the radiance are given
in the space-time domain when the source is punctual and isotropic or 
unidirectional. These analytical
results are compared to Monte-Carlo simulations in four particular situations.
\end{abstract}


\vskip 1em
\maketitle

Transport in disordered and random media is a widely addressed physical
question which plays an important role in several domains of Physics.
Motivated by the kinetic theory of gases, the Boltzmann equation has 
been studied since more than a century.
The radiative transfer equation is a Boltzmann equation
where speed is fixed. It was derived by Chandrasekhar~\cite{chandrasekhar}
to study the radiation transport in a scattering atmosphere. 
Although radiative transfer is mostly used in three-dimensional systems,
the two-dimensional radiative transfer is of interest in several domains, 
such as seismology, where surface waves carry most of the energy.

Some solutions of the two-dimensional radiative transfer equation 
are known analytically. For isotropic scattering and isotropic source, 
the energy distribution has been found by Shang and Gao,
and Sato \cite{shang1988,sato1993} and Paasschens improved these
results by providing the radiance distribution \cite{paasschens1997}.
Recent progress in numerical and analytical solutions have been
made by Liemert and Kienle \cite{liemert2011,liemert2012a}. 
These numerical methods efficiently extend to the three-dimensional
case~\cite{liemert2012b,liemert2014}. The situation in two dimensions
is more favorable to analytical results for many reasons, geometrical
and analytical; let us only mention that the rotation group has a single
parameter and that the Green's function has an algebraic Fourier-Laplace
transform.

Let us write~$q(\vr,\,t,\,\theta)$ the space-time
density of energy flux at~position $\vr$
at the time~$t$ with direction angle~$\theta$. 
$q(\vr,\,t,\,\cdot)$ is called the \emph{radiance} in the
standard terminology in optics. If we integrate the radiance with
respect to~$\theta$, we get the spatial distribution of energy flux
at~$\vr$ and~$t$. In systems
with no absorption, the energy flux distribution integrated over space
is constant and normalized to~$\frac c\ell$ in this work.

The differential equation for $q$ is
\begin{multline}
\derp qt(\vr,\,t,\,\theta)+c\, \uu(\theta)\cdot\nabla q(\vr,\,t,\,\theta)
       +\frac c\ell q(\vr,\,t,\,\theta)=\\
       \frac c\ell\int_{\S1} \varphi\left(\theta-\theta'\right)
           q(\vr,\,t,\,\theta')\d\theta'.
\label{rte}
\end{multline}
The \emph{phase function}~$\varphi$ is an even real valued function 
describing the
distribution of scattering angle. The case where all scattering angles
are equally likely, $\varphi(\theta)=1/2\pi$, 
is the \emph{isotropic} case. 

To solve the equation \eqref{rte} we introduce ``unscattered''
distributions that are the spatial distribution of probability
of particles that have not been scattered. These distributions
are distinguished by a subscript~$_0$. ``Scattered'' distributions
receive the same notations without this subscript.
The probability to meet a scatterer on its trajectory 
at a distance~$r$ from the source is $\e^{-r/\ell}$,
where~$\ell$ is the mean free path. 
The distribution of particles starting from the origin at time $t=0$ 
and moving with speed $c$ with angle~$\theta_0$ that have not been 
scattered at time~$t$ is
\begin{equation}
\fri\left(\vr,\,t,\,\theta_0\right)
  =\frac c\ell \delta^{(2)}\left(\vr-ct\uu(\theta_0)\right)\;\e^{-ct/\ell}.
\label{fri}
\end{equation}
$\fri$ is the 
\emph{unscattered energy distribution from an unidirectional point source}. 
$\delta^{(2)}$ is a two-dimensional Dirac delta function.
In the absence of scattering, the propagation angle~$\theta$ is
preserved, its distribution is a Dirac delta-function
$\delta(\theta-\theta_0)$. This defines the
\emph{unscattered radiance distribution from an unidirectional point source}
\begin{equation}
\fru\left(\vr,\,t,\,\theta;\,\theta_0\right)=
\fri\left(\vr,\,t,\,\theta_0\right) \delta(\theta-\theta_0). 
\label{fru}
\end{equation}
We remark that $\fru\left(\vr,\,t,\,\theta;\,\theta_0\right)$ 
is invariant if one exchanges~$\theta$ and $\theta_0$. As a consequence
$\fri(\vr,\,t,\,\theta)$ is also
the \emph{unscattered radiance from an isotropic source}.
Finally, integrating~$\fri$ over the angle $\theta$, 
we obtain the \emph{unscattered energy distribution from an isotropic source}
\begin{equation}
\fei\left(\vr,\,t\right)=\frac{1}{2\pi}
  \int_{\S 1}\;\fri[](\vr,\,t,\,\theta)\,\d\theta
 =\frac c\ell\frac{\delta(r-ct)}{2\pi r}\e^{-ct/\ell}.
\label{fei}
\end{equation}
The distributions defined by Equations~\eqref{fri}, \eqref{fru} and \eqref{fei}
constitute the building blocks for 
the multiple scattering theory presented in this paper.

\paragraph*{Notations and analytic transforms}
We use the units $c=1$, $\ell=1$. We denote by $\TF f(k)$
the spatial Fourier transform of the radial function $f(r)$ and $\TL f(s)$ 
the time Laplace transform of $f(t)$. We will also use the Hankel
transform as defined in the appendix~\ref{dbleT}.

The Fourier-Laplace transform of $f(r,\,t)$ is denoted by
$\TFL f(k,\,s)$. The leading exponential factor $\e^{-t}$ in 
$\fei[]$, $\fri[]$ and $\fru[]$ results in the shift of the variable $s$
by $1$. All Fourier-Laplace transformed functions carry
this shift and are written only with
their angular dependences, as in 
$\fri[\TFL](\theta)\equiv\fri[\TFL](\vk,\,s-1,\,\theta)$. 
The Fourier-Laplace transforms of the unscattered distributions
admit the following expressions
\begin{eqnarray}
\fei[\TFL]&=&\frac{1}{\sqrt{k^2+s^2}},
\label{fei TFL}\\
\fri[\TFL](\theta)&=&\left(s+\i\vk\cdot\uu(\theta)\right)^{-1},
\label{fri TFL}\\
\fru[\TFL](\theta;\,\theta_0)&=&\delta\left(\theta-\theta_0\right)
  \fri[\TFL](\theta_0).
\label{fru TFL}
\end{eqnarray}

\section{Analytical derivation}
The energy distribution $\gei$ of the two-dimensional 
isotropic radiative transfer has been provided by 
Gao \& Shang~\cite{shang1988}, Sato~\cite{sato1993} and 
Paasschens~\cite{paasschens1997}, the latter having also 
given the radiance solution~$\gri$. 
This section is dedicated to the analytical computation of 
the \emph{scattered radiance from an unidirectional point source} $\gru$.
On the way to this result we compute the scattered Green's
functions~$\gei$ and $\gri$. 

\subsection{Solutions in the Fourier-Laplace domain}
The radiative transfer equation~\eqref{rte} governing 
$\gri$ in the Fourier-Laplace domain 
rewrites for isotropic scattering
\begin{equation}
\left(s+\i\vk\cdot\uu(\theta)\right)\gri[\TFL](\theta)
 =\gri[\TF](\vk,\,0,\,\theta) + 
  \frac{1}{2\pi}\int_{\S1}\gri[\TFL](\theta')\d\theta'.
\label{rtei}
\end{equation}
The initial condition 
$\gri(\vr,\,0,\,\theta)=\delta^{(2)}(\vr)$
enters into the equation \eqref{rtei} as
$\gri[\TF](\vk,\,0,\,\theta)=1$.
We multiply the equation \eqref{rtei}
by $\fri[\TFL](\theta)$ as given by~\eqref{fri TFL} and we obtain
\begin{eqnarray}
\gri[\TFL](\theta)
&=&\fri[\TFL](\theta)+\fri[\TFL](\theta)\frac1{2\pi}\int_{\S1}\gri[\TFL](\theta')\d\theta'
\nonumber \\
&=&\fri[\TFL](\theta)+\fri[\TFL](\theta)\,\gei[\TFL].
\label{gri Dyson}
\end{eqnarray}
We notice that~$\gri[\TFL]$ will be known as soon as we know~$\gei[\TFL]$. 
From the integration of equation \eqref{gri Dyson} over~$\theta$
we get, using the definition~\eqref{fei}, the Green-Dyson 
relation
\begin{equation}
\gei[\TFL]=\gei[\TFL]_0+\gei[\TFL]_0\, \gei[\TFL]
\label{gei Dyson}
\end{equation}
and deduce from the expression of~$\fei[\TFL]$ in equation~\eqref{fei TFL}
the \emph{scattered energy distribution from an isotropic source}
\begin{equation}
\gei[\TFL]=\frac{\fei[\TFL]}{1-\fei[\TFL]}
=\fei[\TFL]+\fei[\TFL]^2+\fei[\TFL]^3+\cdots
=\frac{1}{\sqrt{s^2+k^2}-1}. 
\label{gei TFL}
\end{equation}
Each order of the expansion corresponds to a given number of scattering events
the particle has experienced. 
From the expression~\eqref{gri Dyson} we find the
\emph{scattered radiance distribution from an isotropic source} to be
\begin{equation}
\gri[\TFL](\theta)=\frac{\fri[\TFL](\theta)}{1-\fei[\TFL]}=
  \fri[\TFL](\theta)+\fri[\TFL](\theta)\fei[\TFL]
 +\fri[\TFL](\theta)\fei[\TFL]^2+\cdots
\label{gri TFL}
\end{equation}
in which we can se that only the last scattering event depends on the angle. 
Conversely, the
\emph{scattered energy distribution from an unidirectional point source}
is given by the same formula written as
\begin{equation*}
\gri[\TFL](\theta) =\fri[\TFL](\theta)
+\fei[\TFL]\fri[\TFL](\theta)+\fei[\TFL]^2\fri[\TFL](\theta)+\cdots
\end{equation*}
in which direction is lost after the first scattering event.
If the source is unidirectional, the \emph{scattered radiance distribution 
from an unidirectional point source}
is therefore given by the relation
\begin{equation}
\gru[\TFL](\theta;\,\theta_0)
=\fru[\TFL](\theta;\,\theta_0)+\fri[\TFL](\theta)\gri[\TFL](\theta_0).
\label{gru TFL}
\end{equation}
We can now use the solutions~\eqref{gei TFL}, \eqref{gri TFL} 
and \eqref{gru TFL} to 
give the expressions of this functions in the space-time domain.

\subsection{Solutions in the space-time domain}
To find the expression of $\gei[]$ in the space-time domain
we use the simultaneous Hankel-Laplace inverse transform of order zero
(see the appendix) with the function
$\TL f(s)=s/(s-1)$ (and thus~$f(t)=\delta(t)+\e^t\Theta(t)$) and we obtain
\begin{equation}
 \gei[](r,\,t)=c\frac{\e^{-ct/\ell}}{2\pi\ell r}\delta(ct-r)+
   \frac{\e^{-ct/\ell+cT/\ell}}{2\pi\ell^2 T}\Theta(ct-r),
\label{gei}
\end{equation}
the energy distribution from an isotropic source as already found by Shang 
and Gao and by Sato. 
We have used the notation $T=\sqrt{t^2-r^2/c^2}$.
Note that the simultaneous inverse transform
was performed thanks to the fact that $\gei[\TFL]$ is a 
function of $\fei[\TFL]^{-1}$.
For $t\gg r$, this solution approaches the Gaussian distribution
of diffusion, with $D=\frac12=c\ell/2$.
To compute $\gri$ we can use the space and time convolution
defined by the Equation~\eqref{gri Dyson} which yields
the expression 
\begin{equation}
\gri(\vr,\,t,\,\theta)=\fri(\vr,\,t,\,\theta)
   +\frac{\e^{-ct/\ell}}{2\pi\ell^2}\frac{\e^{cT/\ell}}
    {t-\vr\cdot\uu(\theta)/c}\Theta(ct-r).
\label{gri}
\end{equation}

\begin{figure}
\centering
\includegraphics[width=0.45\textwidth]{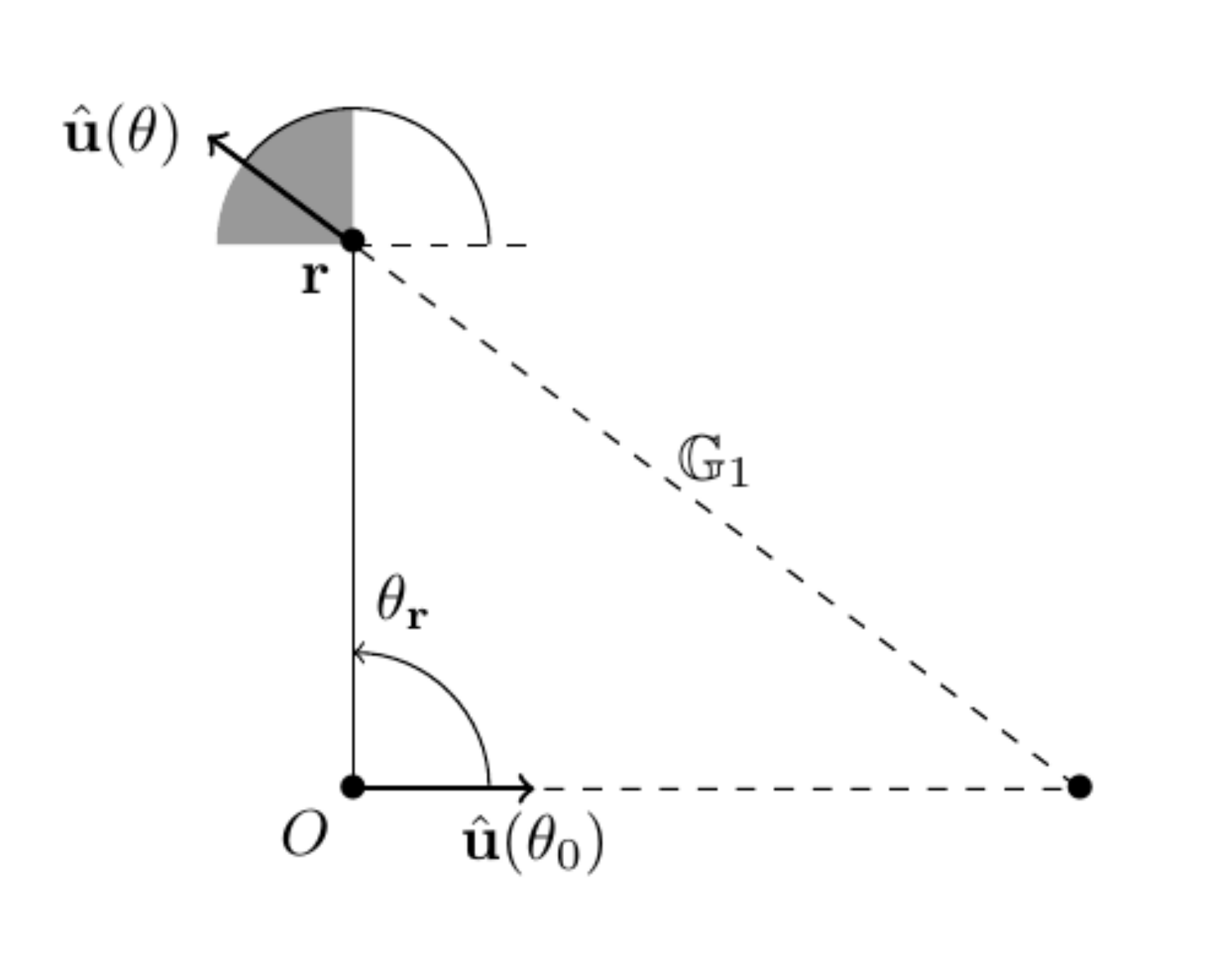}
\caption{\label{fig:G1}
Description of the geometric variables in the situation where
$\vr$ and $\theta_0$ are fixed. The trajectory with a
single scattering event is drawn as a dashed line. Around the point $\vr$, the
shaded region shows the angle interval in which $\gsru$ contributes to the
intensity.}
\end{figure}

The expression \eqref{gri} 
was first derived by Paasschens~\cite{paasschens1997}.
Our work extends his results to the radiance distribution from
an unidirectional point source, $\gru$. 
To compute~$\gru$, we use the result~\eqref{gri}
together with the relation~\eqref{gru TFL} 
which defines a convolution in the space-time domain. 
After integration with respect to the space coordinate we get
\begin{widetext}
\begin{equation}
\gru(\vr,\,t,\,\theta;\,\theta_0)= 
     \fru(\vr,\,t,\,\theta;\,\theta_0)+
     \gsru(\vr,\,t,\,\theta;\,\theta_0)+
     \Theta(t-r)\frac{\e^{-t}}{2\pi}
     \int_0^{\frac{t^2-r^2}{2\left(t-\vr\cdot\uu(\theta_0)\right)}}
     \frac{\e^{\sqrt{t^2-\vr^2-2\tau\left(t-\vr\cdot\uu(\theta_0)\right)}}}
          {t-\tau-\vr\cdot\uu(\theta)+\tau\uu(\theta)\cdot\uu(\theta_0)}
     \d\tau,\nonumber
\end{equation}
\end{widetext}
where $\gsru$ is the single scattering contribution arising from
the convolution of $\fri$ with itself (see the figure~\ref{fig:G1}
and the appendix \ref{sec:gsru}).

If $\vr=t\uu(\theta_0)$, 
the integral vanishes (all the energy is contained in the
ballistic term $\fru$), otherwise we can perform
the change of variable~$2(t-\vr\cdot\uu(\theta_0))\tau=t^2-r^2-y^2$
and we get for~$\theta\neq\theta_0$
\begin{equation*} 
  \gru=\fru+\gsru+
   \frac{\e^{-ct/\ell}}{2\pi}\frac{2\Theta(t-r)}{1-\cos(\theta-\theta_0)}
     \int_0^T\frac{y\,\e^{y}}{X^2+y^2}\d y,
\end{equation*}
where $X$ is defined by
\begin{equation}
X=\frac1\ell\sqrt{\frac{2\left(ct-\vr\cdot\uu(\theta)\right)
               \left(ct-\vr\cdot\uu(\theta_0)\right)}
                {1-\cos(\theta-\theta_0)}       -c^2T^2}.
\label{rho}
\end{equation}
We finally obtain our main result for isotropic 
scattering (with $\theta\neq\theta_0$)
\begin{widetext}
\begin{eqnarray}
\gru(\vr,\,t,\,\theta;\,\theta_0)=
\fru(\vr,\,t,\,\theta;\,\theta_0)+\gsru(\vr,\,t,\,\theta;\,\theta_0)+
   c\frac{\e^{-ct/\ell}}{2\pi\ell^3}
   \frac{2\Theta(ct-r)}{1-\cos(\theta-\theta_0)}\Re\left[
 \EE\left(\i X\right)\e^{\i X}-\EE\left(\i X-\frac c\ell T\right)\e^{\i X}
   \right].
\label{gru}
\end{eqnarray}
The function $\EE[n]$  
is the $n^{\rm th}$ order exponential integral function as defined
in \cite[chap. 5]{abramowitzstegun}. The expression~\eqref{gru}
has been obtained using the antiderivative 5.1.44 in this reference.
In the case where~$\theta=\theta_0$, the integral gives the result
\begin{equation}
 \gru(\vr,\,t,\theta_0;\,\theta_0)=\fri(\vr,\,t,\,\theta_0)+
  \Theta(ct-r)\,c
  \frac{\e^{-ct/\ell}}{4\pi\ell}\;
  \frac{1+\left(c T/\ell-1\right)
      \e^{c T/\ell}}
       {\left(ct-\vr\cdot\uu(\theta_0)\right)^2}.
\label{gru'}
\end{equation}
\end{widetext}
The term $\fri$ is the unscattered contribution while
the second term is a scattered contribution of second order 
(at least two scattering events have occured). There are no single scattering
contributions from~$\gsru$ in \eqref{gru'}.

\subsection{Steady-state solutions}
The time-dependent scattered solutions 
measured at a given point~$\vr$ exhibit a variety of behaviours 
that can be exploited when using pulse sources. However, some experimental
setups may require the use of a steady source. Hence, we discuss here
the steady-state
solutions of the radiative transfer equation in two dimensions.
We have to first remark that the large time regime
is diffusive and as Brownian motion in two dimensions is recurrent, a
steady source would yield a diverging energy density as time goes to infinity.
However, in the presence of an absorption rate~$\mu>0$, all 
unscattered and scattered
Green's functions get a leading regularizing factor $\e^{-\mu t}$. Such a
constant rate could come from energy dissipation under another form (like,
typically, heat) or account for losses into the third dimension.
Since the dimension two is the critical dimension for Brownian recurrence,
we expect the steady-state distribution to diverge logarithmically as
$\mu$ or $r$ goes to zero.

In the presence of absorption, the steady-state 
counterpart~$f\stst(r)$
of a Green's function~$f(r,t)$ is well defined and we have
\begin{equation}
 f\stst(r)=\int_0^\infty \e^{-\mu t}f(r,t)\d t=\TL f(r,\mu).
\label{intt}
\end{equation}
It could also be obtained as the inverse Fourier transform of 
$\TF{f}\stst(k)=\TFL f(k,\mu)$. Denoting by $\alpha=\ell^{-1}+\mu/c$
the total extinction rate, the unscattered energy distribution
is $\fei\stst(r)=\e^{-\alpha r}/(2\pi\ell r)$ and
we show in the appendix~\ref{sec:stst} that $\gei\stst=\fei\stst+\geistst$ where
\begin{equation}
 \geistst(r)
  = \frac{1}{2\pi}
  \sum_{n=0}^\infty
  \frac{\left(-\kappa r\right)^n}{n!}
      \EE[n+1]\left(\frac{\mu r}{2c}\right).
\label{gss}
\end{equation}
with $\kappa=\ell^{-1}+\mu/2c$.  
The expression~\eqref{gss} is exact and is convenient for small~$r$
expansion, where it converges quickly. Using the steepest descent
method, we obtain an approximation 
of $\gei\stst(r)$ for large~$r$ ($r\gg c/\mu$) as
\[\geistst(r)\mathop{\approx}_{r\to\infty}
\frac{1}{(2\pi)^{3/2}\ell^2}
\frac{\e^{-r\sqrt{2\kappa\mu/c}}}{\sqrt{r(2\kappa\mu/c)^{1/2}}}.\]
For large $\mu$ ($\mu\gg c/\ell$) we find $\geistst(r)\approx
\mathrm{K}_0(\mu r)/(2\pi\ell^2)$. 
In both cases, 
we observe a slower energy decay away from the source than for the 
unscattered energy distribution.

The unscattered radiance distributions are proportional to $\fei\stst$. 
We easily find 
$\fri\stst(\vr,\theta)=\fei\stst(r) \delta(\theta-\theta_\vr)$ and
$\fru\stst(\vr,\theta;\theta_0)
=\fri\stst(\vr,\theta)\delta(\theta-\theta_0)$. 
Since the equation \eqref{gri Dyson} states that
$\gri[\TFL]=\fri[\TFL]+\fri[\TFL]\gei[\TFL]$
we have 
$\gri[\TF]\stst=\fri[\TF]\stst
+\fri[\TF]\stst\gei[\TF]\stst$.
The steady-state distributions have the same convolution relations
as the time dependent ones. The distribution
$\gri\stst$ 
cannot be computed exactly, but we should remark that 
near the source, the lowest order of scattering dominates the
distribution. The unscattered term, proportional to 
$\fei\stst$, dominates everywhere it is
not equal to zero. A single scattering contribution appears in
$\gri\stst$, we denote it by $\gsristst$.
We can therefore decompose $\gri\stst$ into
$\gri\stst=\fri\stst+\gsristst+\gristst$.
On the figure~\ref{fig:G1}, the shaded region corresponds to the
geometric configuration where the radiance distribution from an unidirectional
point source has a contribution from single scattering. 
If single scattering
does not contribute, the main contribution is from double scattering,
$\gdrustst$ (we do not provide an expression for this contribution).
The distribution $\gru\stst$ decomposes into 
$\fru\stst+\gsru\stst+\gdrustst+\grustst$. 
The distributions $\gsristst$ and $\gsru\stst$ are given in the appendix~\ref{sec:gsru}.

\section{Numerical simulations}
We compare the solution \eqref{gru} to statistics obtained from
a Monte-Carlo simulation of the two dimensional isotropic Boltzmann equation.
We start random walks from the origin at $t=0$ with propagation
angle~$\theta_0=0$.  The step length~$x$ is distributed exponentially according
to the probability distribution $p(x)=\e^{-x}$. After each step, we chose
a random angle~$0\leq\theta<2\pi$ for the propagation.

\begin{figure}
\centering
\includegraphics[width=0.45\textwidth]{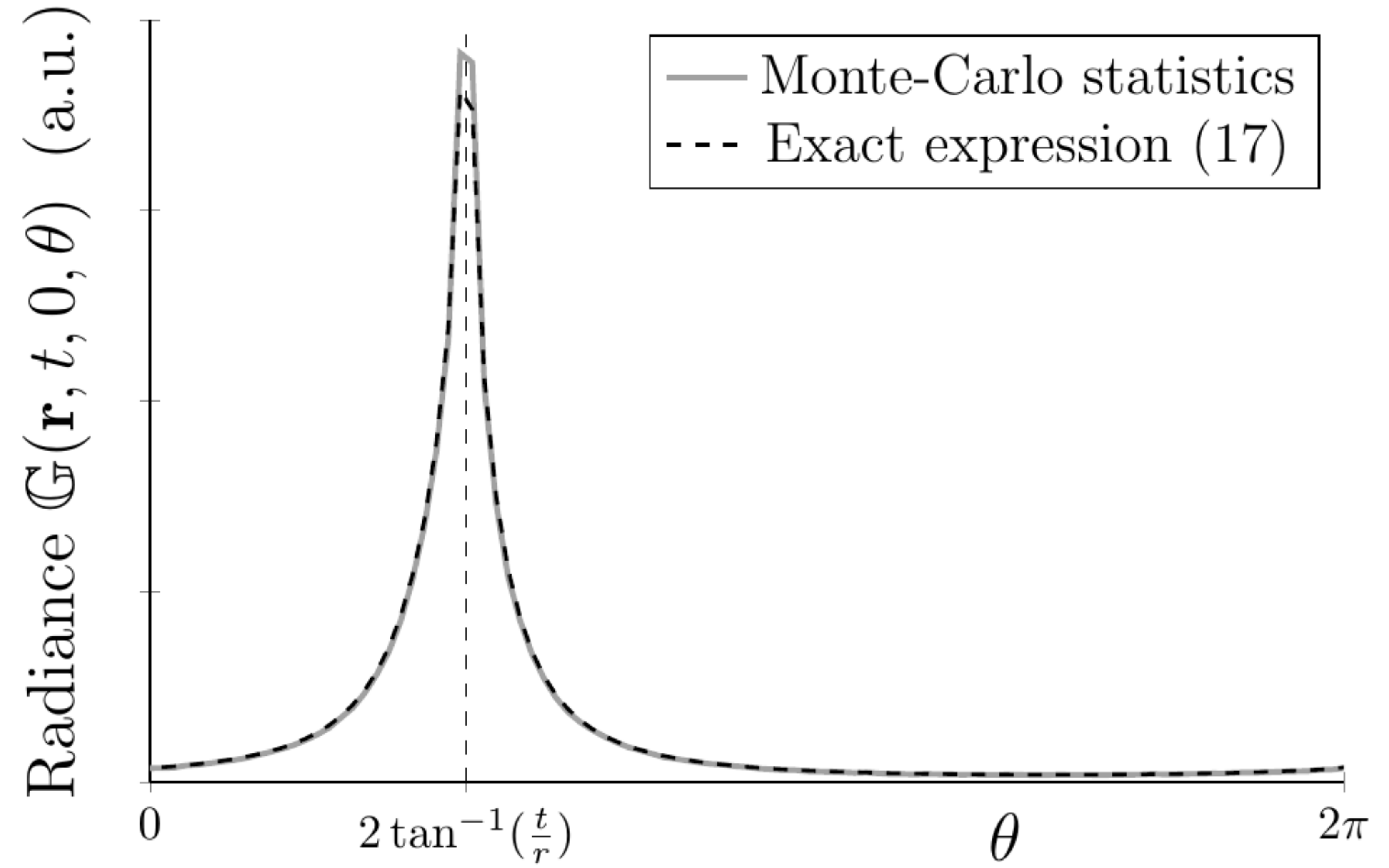}
\caption{\label{fig:theta}
Radiance for fixed $\theta_0=0$, $t=1.1$ 
and $\vr=(0,1)$ (thus $\theta_\vr=\frac\pi2$)
as a function
of the propagation angle~$\theta$. The Monte-Carlo simulations have been
performed until $10^7$ trajectories with at least two scattering events
are found satisfying the fixed conditions for $t$ and $r$
within $\Delta t=0.05$ and $\Delta r=0.05$.
A total of $7.16\times10^{10}$ trajectories have been computed.}
\end{figure}

\begin{figure}
\centering
\includegraphics[width=0.45\textwidth]{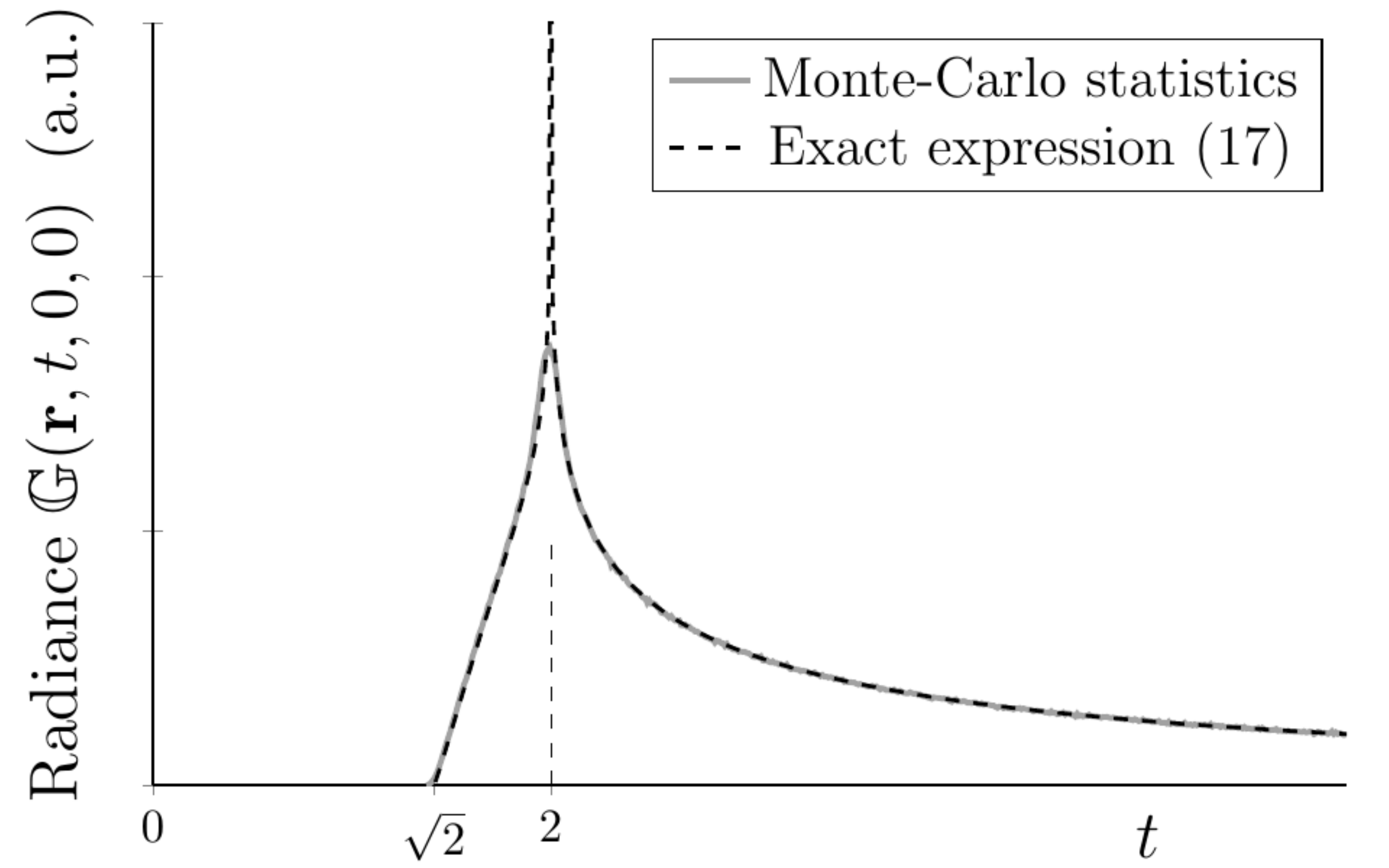}
\caption{\label{fig:t}
Radiance in the direction orthogonal to the initial propagation
direction as a function of time at the position $\vr=(1,1)$. 
Single scattering contributions have been removed. The Monte-Carlo 
simulations have been
performed until $10^7$ trajectories with at least two scattering events
are found satisfying the fixed conditions for $\vr$ and $\theta$
within $\Delta r=0.05$ and $\Delta\theta=\frac\pi{100}$.
A total of $2.85\times10^{10}$ trajectories have been computed.}
\end{figure}
The figure~\ref{fig:G1} shows a situation where~$\theta_\vr=\frac\pi2$ used
in the Monte-Carlo simulations. 
If the random walker approaches the ``target'' position~$\vr$ by a 
distance less than $\Delta r$ between the times $t-\Delta t/2$
and $t+\Delta t/2$, we store the value taken by $\theta$ during
the corresponding step. The statistics
of $\theta$ follow the distribution 
$\gru(\vr,\,t,\,\theta,\,0)\Delta t\times \pi(\Delta r)^2$. 
The distribution exhibits a peak at $\theta=2\tan^{-1}(\frac tr)$, 
corresponding to the single scattering trajectory. 
The results of these simulations is 
displayed in the figure~\ref{fig:theta}, they compare the distributions of 
the propagation angle~$\theta$ for fixed position~$\vr$ and time~$t$
obtained by Monte-Carlo simulations to the predicted formula~\eqref{gru}.

The figure~\ref{fig:t} displays the radiance
at angle $\theta=\pi/2$ as a function of time at the
fixed point $\vr=(1,1)$.
If the random walker approaches the ``target'' position~$\vr$ by a
distance less than $\Delta r$ and the propagation angle $\theta'$ is such
that $\cos(\theta'-\theta)>\cos(\Delta\theta)$, we store the value of $t$
corresponding to the closest point along the matching step. 

\begin{figure}
\centering
\includegraphics[width=0.45\textwidth]{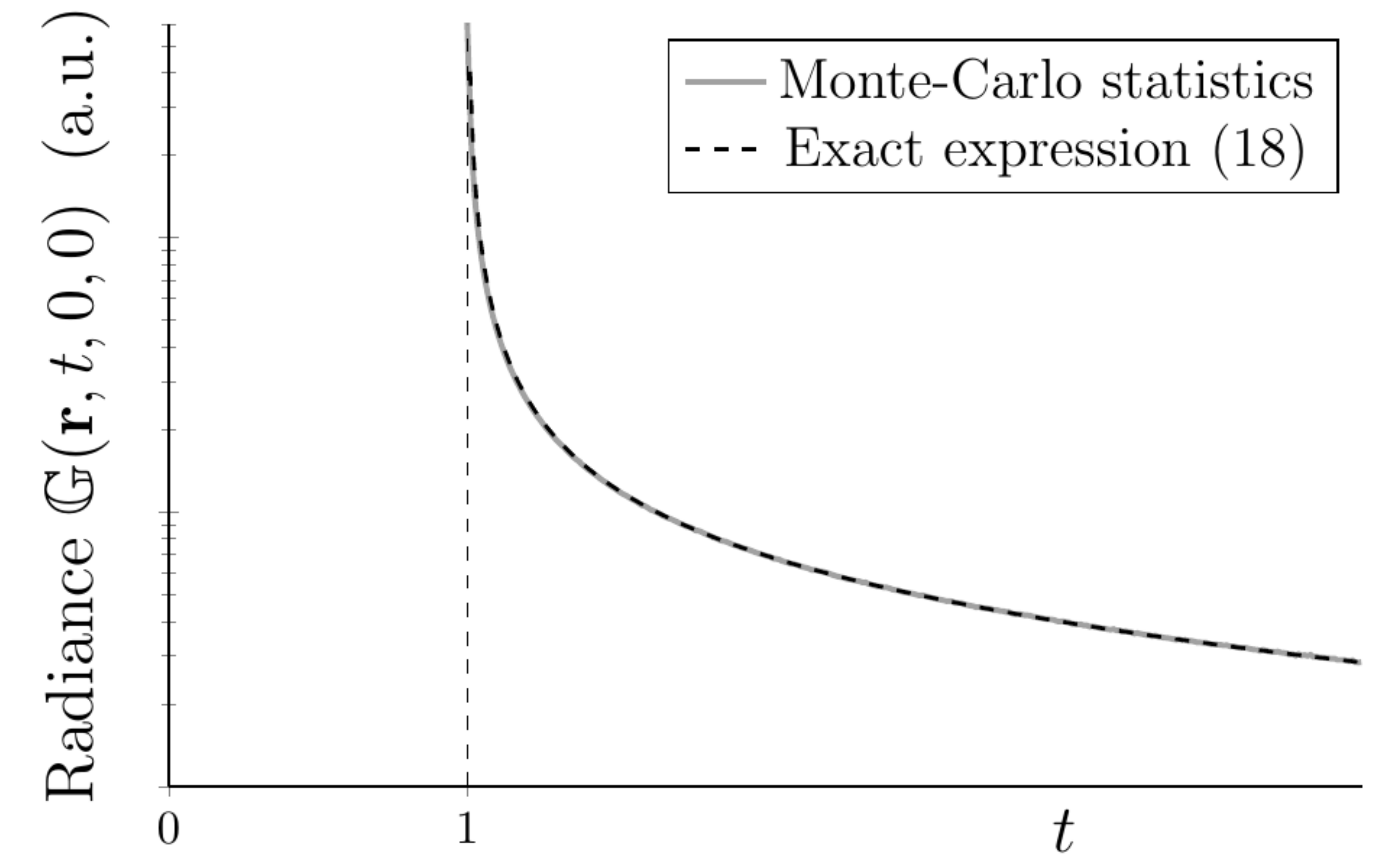}
\caption{\label{fig:t axe}Radiance in the initial direction
of propagation as a function of time at the position $\vr=(1,0)$.
Non-scattered contributions have been removed. The Monte-Carlo simulations
have been performed until $10^8$ trajectories are found satisfying the 
fixed conditions
for $\vr$ and $\theta$ within $\Delta r=0.05$ and $\Delta\theta=\frac\pi{100}$.
A total of $1.49\times 10^{11}$ trajectories have been computed.}
\end{figure}

\begin{figure}
\centering
\includegraphics[width=0.45\textwidth]{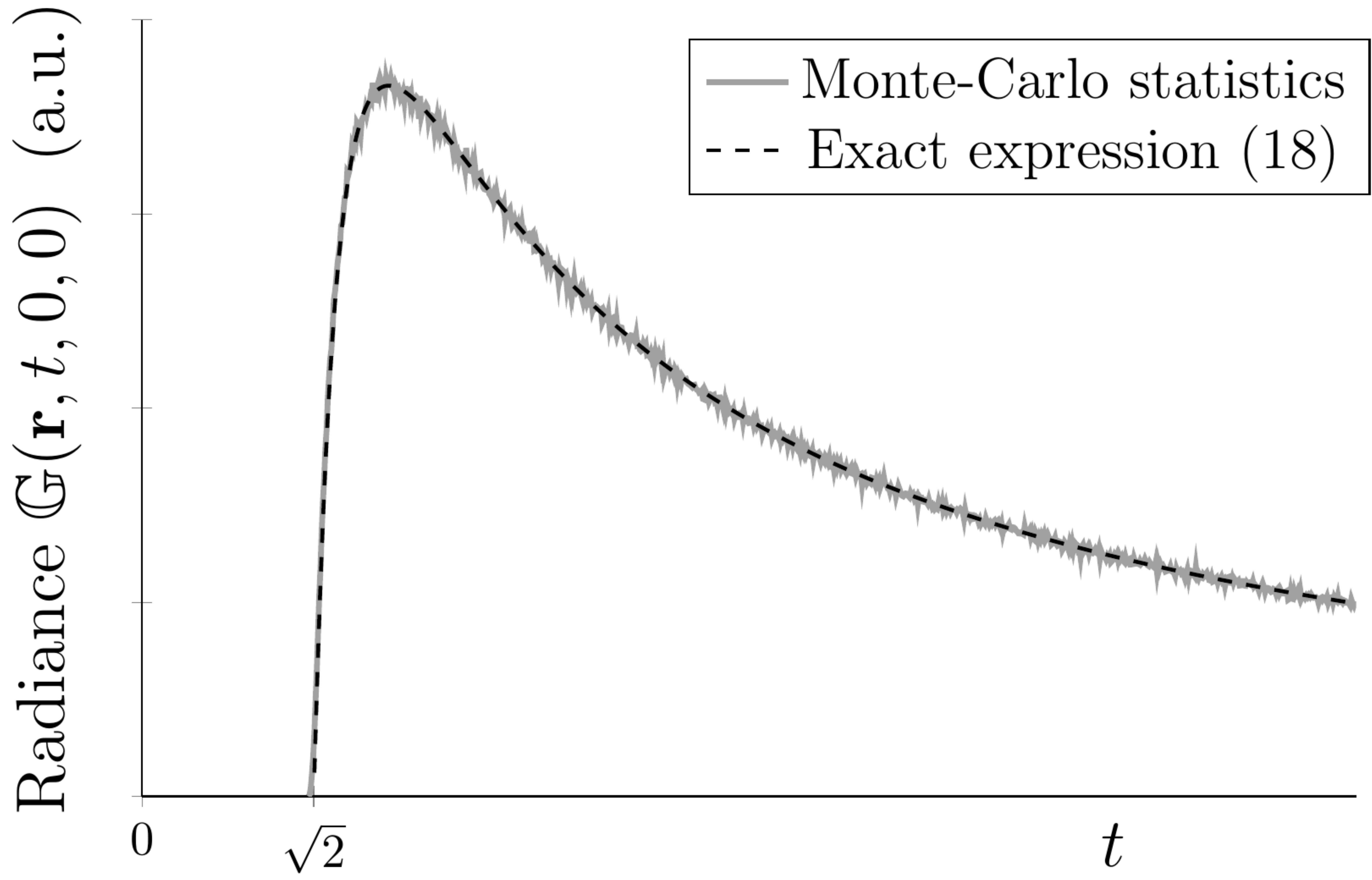}
\caption{\label{fig:t off}Radiance in the initial direction
of propagation as a function of time at the position $\vr=(1,1)$.
Non-scattered contributions have been removed. The Monte-Carlo simulations
have been performed until $10^8$ trajectories are found satisfying the 
fixed conditions
for $\vr$ and $\theta$ within $\Delta r=0.05$ and $\Delta\theta=\frac\pi{100}$.
A total of $3.77\times 10^{11}$ trajectories have been computed.}
\end{figure}

The figures~\ref{fig:t axe} and~\ref{fig:t off} show the radiance
at angle $\theta=\theta_0=0$, the pathological case where Equation~\eqref{gru}
has to be replaced by~\eqref{gsru}, as a function of time at the
fixed points $\vr=(1,0)$ and $\vr=(1,1)$ respectively. The method is the same
as explained for figure~\ref{fig:t}.

In the figures, the normalization of the numerical distributions has been
adjusted, no other parameters have been tuned.

\section{Outlook}
We have computed the exact solutions of the radiative transfer
equation in two
dimensions with angular resolution both at the source and the
receiver. The time-dependent solutions are useful
for the signal analysis when the source is modulated.
The steady-state solutions could only be estimated: The approximations
we have obtained are asymptotically close to the exact solutions when
the absorption is strong or when the distance from the source is large.
When absorption is low and the distance from the source is small, the
single scattering contribution grows logarithmically and dominates the
radiance. 

We now  briefly discuss the applications of these results.
The angular resolution of the theoretical radiance will be useful for
analysing data collected with full or partial angular dependences. 
Angular dependences are easily acessible
in optics: Some light sources, such as laser beams, are inherently
unidirectional, and collimated receivers can be used to measure the
radiance. In acoustics, the so-called beam-forming methods
use the signals recorded by an array of aligned receivers
to select the sound from an incoming direction.
This technique also works with an arrays of sources
to produce a unidirectional source of sound. 
These beam-forming techniques are also 
frequently used in geophysics as well, with seismic waves. 
In these fields, using the angular dependences of scattered waves
would represent a substantial increase of the available amount of data.
We expect that such an increase will help improving imaging methods.

\appendix
\section{Simultaneous Hankel-Laplace transform}
\label{dbleT}
We give here a proof of the simultaneous Hankel-Laplace transformation
formula of arbitrary order~$n$. This transformation is more general than the
case $n=0$ used in the main text.
The two-dimensional Fourier transform of a function 
$f(\vr)=\sum_{n\in\mathbb Z}f_n(r)\e^{\i n\theta_\vr}$
is by definition $\sum_{n} \,\i^n\e^{\i n\theta_\vk}\TH f_{n}(k)$, with
$\TH f_n$ the Hankel tranform of  
$\left| n\right|^\textrm{\scriptsize th}$ order of~$f_{n}$. 

The $n^\textrm{\scriptsize th}$ order Hankel transform of 
$\delta(r-x)/2\pi r$ is 
given by $J_n(kx)$. Replacing $x$ by $\sqrt{t^2-u^2}$ and multiplying
by $(t-u)^{n/2}(t+u)^{-n/2}$ 
we recognize the Laplace transform 29.3.97 in
\cite{abramowitzstegun} which is equal to
$\exp\left(-u\sqrt{s^2+k^2}\right)\xi^n$ with 
\begin{math}
\xi=k/(s+\sqrt{s^2+k^2})
\end{math}.
The expression $\exp\left(-u\sqrt{s^2+k^2}\right)\xi^n$ is therefore the
$n$-Hankel-Laplace transform of 
$(t-u)^{n/2}(t+u)^{-n/2}\delta(r-\sqrt{t^2-u^2})/2\pi r$.
Multiplying both these expressions by an arbitrary function 
$f(u)$ and integrating from $u=0$ to infinity
we find that
\begin{equation}
\phantom{\frac1{2\pi}}\frac{1}{\sqrt{s^2+k^2}}\left(\frac{k}{s+\sqrt{s^2+k^2}}\right)^n
\TL f\left(\sqrt{s^2+k^2}\right)
\label{THL}
\end{equation}
is the $n^{\textrm{th}}$ order Hankel-Laplace transform of
\begin{equation}
\frac{1}{2\pi}\frac{\Theta(t-r)}{\sqrt{t^2- r^2}}
\left(\frac{r}{t+\sqrt{t^2-r^2}}\right)^n
f\left(\sqrt{t^2-r^2}\right).
\label{f}
\end{equation}
We call the transformation between \eqref{THL} and \eqref{f} a
simultaneous double transform. Simultaneous double transforms
have been introduced recently in Ref. \cite{rossetto2013b}.
It should be noticed that the simultaneous inversion for anisotropic
scattering is possible because the expansion of $\grua[\TFL]$ 
is a rational function of $\fei[\TFL]^{-1}=\sqrt{k^2+s^2}$.
\\[0.5em]

\section{The single scattering functions}
\label{sec:gsru}
The term $\gsru$ in equation \eqref{gru} is purely geometric. Its
computation is straighforward and yields
\begin{multline}
 \gsru(\vr,\,t,\,\theta;\,\theta_0)=
  \e^{-ct/\ell} \\
  \delta^{(2)}\left[
    \big(\vr-ct\uu(\theta)\big)\times\big(\uu(\theta_0)
    -\uu(\theta)\big)\right]\\
       \Theta\left[\big(\vr-ct\uu(\theta)\big)\cdot
        \big(\uu(\theta_0)-\uu(\theta)\big)\right]\\
  \Theta\left[ct\big(1-\cos(\theta-\theta_0)\big)-
    \vr\cdot\big(\uu(\theta)-\uu(\theta_0)\big)\right].
\label{gsru}
\end{multline}
In the steady state regime, it reduces to
\begin{multline}
\gsru\stst(\vr,\,\theta;\,\theta_0)=\\
  \frac{
  \Theta\left[\cos(\theta_\vr-\theta_0)-
     \cos(\theta_\vr-\theta)\cos(\theta-\theta_0)\right]}
   {4\pi^2\ell^2\; |\sin(\theta-\theta_0)|} \\
  \exp\left(-\alpha r\left|
    \frac{\sin(\theta_\vr-\theta)-\sin(\theta_\vr-\theta_0)}
   {\sin(\theta-\theta_0)}\right|\right). 
\end{multline}
In the steady-state regime, the contribution of single scattering
to the radiance distribution from an isotropic source arises from
the convolution of $\fei\stst$ and $\fri\stst$, it is equal to
\begin{multline}
\gsristst(\vr,\,\theta)=\frac{1}{4\pi^2\ell^2}
  \exp\left(-\alpha r\cos(\theta_\vr-\theta)\Big)\right)\\
  \EE\left(\alpha r
           \Big(1-\cos(\theta_\vr-\theta)\Big)\right).
\end{multline}

\section{Steady-state approximations}
\label{sec:stst}
The steady-states solutions and approximations are based on the following
expressions. 
We perform the change of variable 
$t=\frac{r}{2c}\left(x+\frac1x\right)$ in the time integral \eqref{intt}
of $\gei$ given by equation \eqref{gei} and we get
\[ \geistst(r)=\frac{1}{2\pi}
 \int_1^\infty\exp\left(-\frac{\mu r}{2}x-\frac{\kappa r}{x}\right)\frac{\d x}x.\]
We obtain~the equation \eqref{gss} 
by expanding the exponential $\e^{-\kappa r/x}$ into
the series $\sum_n \frac{(-\kappa r/x)^n}{n!}$. The large~$r$ approximation
is obtained by second order polynomial expansion of 
$\frac{\mu r}{2}x+\frac{\kappa r}{x}$ 
around the minimum at $x=\sqrt{2\kappa/\mu}$.

\vskip 1em
\providecommand{\bysame}{\leavevmode\hbox to3em{\hrulefill}\thinspace}
\providecommand{\MR}{\relax\ifhmode\unskip\space\fi MR }
\providecommand{\MRhref}[2]{%
  \href{http://www.ams.org/mathscinet-getitem?mr=#1}{#2}
}
\providecommand{\href}[2]{#2}


\end{document}